\documentstyle[12pt]{article}

\begin{document}

\thispagestyle{empty}

\hfill \parbox{45mm}{{ECT*-98-002} \par March 1998} 

\vspace*{15mm}

\begin{center}
{\LARGE ``Tunneling'' Amplitudes of a Massless Quantum Field.}

\vspace{22mm}

{\large Giovanni Modanese}%
\footnote{e-mail: modanese@science.unitn.it}

\medskip

{\em European Centre for Theoretical Studies in Nuclear Physics and 
Related Areas \par 
Villa Tambosi, Strada delle Tabarelle 286 \par
I-38050 Villazzano (TN) - Italy}

\bigskip \bigskip

\medskip

\end{center}

\vspace*{10mm}

\begin{abstract}

We propose a method for the approximate computation of the Green
function of a scalar massless field subjected to potential barriers of 
given size and shape in spacetime. The potential of the barriers has the form
$V(\phi)=\xi (\phi^2-\phi_0^2)^2$; $\xi$ is very large and $\phi_0$ 
very close to zero, the product $\xi \phi_0^2$ being finite and small. 
This is equivalent to the insertion of a suitable constraint in the 
functional integral for $\phi$. The Green function contains a
double Fourier transform of the characteristic function of the region 
where the potential has support.

\medskip
\noindent
03.70.+k Theory of quantized fields.

\noindent
31.15.Kb Path-integral methods.

\bigskip 

\end{abstract}

Several problems in Quantum Field Theory require the introduction of
constraints or boundary conditions on the field. The best known case
is the Casimir effect [1]. 

\newpage

Usually problems of this kind are reduced to partial differential
equations for the propagator of the field (and thus to eigenvalues
problems), with suitable boundary conditions [2]. We would like to show
that in certain cases it is possible to obtain, to lowest order at
least, an explicit solution, that can be evaluated numerically. 

Let us consider a massless scalar field $\phi$, coupled to an external source
$J$. The system is described by an Euclidean functional integral, of
the form
\begin{equation}
W[J]=\int d[\phi] \exp \left\{-\int d^4x \, (\partial \phi)^2 - S'[\phi,J]
\right\}
\label{e1}
\end{equation}

Now we want to impose on the field the condition
\begin{equation}
\phi^2(x)=\phi_0^2 \qquad {\rm for} \qquad x \, \epsilon 
\, \Omega_i, \ i=1...n,
\label{e3}
\end{equation}
where the $\Omega_i$'s are regions in spacetime. In order to fix the ideas,
we can suppose that the regions $\Omega_i$ represent compact, distinct
portions of space, and are thus infinitely extended in the temporal
direction. In an electrostatic analogy, they could represent perfect
conductors of arbitrary shape, fixed at certain positions in space.

In order to implement condition (\ref{e3}), let us choose the
source term
\begin{equation}
S'[\phi,J]=\xi \int d^4x \, J_\Omega(x) \left[ \phi^2(x)-\phi_0^2 \right]^2,
\label{e4}
\end{equation}
where $J_\Omega(x)=1$ for $x \, \epsilon \, \Omega_i$, $J_\Omega(x)=0$
elsewhere. In the limit $\xi \to \infty$, the exponential in 
(\ref{e4}) acts in the functional integration (\ref{e1}) like a
$\delta$-function and the square of the field, $\phi^2(x)$, is forced
to assume in each configuration the value $\phi_0^2$ within the
regions $\Omega_i$.

Expanding the square in (\ref{e4}) and inserting $S'$ into (\ref{e1}) we 
obtain for $W[J]$ the expression
\begin{equation}
W[J]=\int d[\phi] \exp \left\{ -\int d^4x \, \left[(\partial \phi)^2 - 
2\xi \phi_0^2 J_\Omega(x) \phi^2(x)+ \xi J_\Omega(x) \phi^4(x)
+ \xi J_\Omega(x) \phi_0^4 \right] \right\}.
\end{equation}
The last term in the square bracket is constant with respect to $\phi(x)$,
and its exponential can be factorized out of the functional integral.
In a first instance -- for weak
fields -- we can disregard the $\phi^4(x)$ term. We are then led to 
consider a 
quadratic functional integral, and the "modified propagator"
$G(x,y)=\langle \phi(x) \phi(y) \rangle_J$, which by definition
satisfies the equation
\begin{equation}
\left[\partial^2_x + \gamma J_\Omega(x)\right] G(x,y) = -(2\pi)^4 
\delta^4(x-y), \label{e6}
\end{equation}
where $\gamma=2\xi \phi_0^2 >0$. Let us focus on the case when
$\phi_0^2=0$ inside the regions $\Omega_i$ and let us take the limit
$\phi_0 \to 0$ and $\xi \to \infty$ in such a way that $\gamma$ is finite
and very small, so that the term $\gamma J_\Omega(x)$ in eq.\ (\ref{e6})
constitutes only a small perturbation, compared to the kinetic term.
Then we can set
\begin{equation}
G(x,y)=G^0(x,y)+\gamma G'(x,y),
\end{equation}
where $G^0(x,y)$ is the propagator of the free scalar field, and we find
immediately that $G'(x,y)$ satisfies the equation
\begin{equation}
\partial^2_x G'(x,y) = - J_\Omega(x) G^0(x,y).
\label{e8}
\end{equation}

In general $G'(x,y)$, unlike $G^0(x,y)$, will not depend only on $(x-y)$,
because the source breaks the translation invariance of the system.
It will then be necessary, in order to go to momentum space, to consider
the Fourier transform of $G'(x,y)$ with respect to both arguments.
We define $\tilde{G}'(p,k)$ and $\tilde{J}_\Omega(p)$ as follows:
\begin{equation}
G'(x,y)=\int d^4p \int d^4k \, e^{ipx} e^{iky} \tilde{G}'(p,k)
\end{equation}
and
\begin{equation}
J_\Omega(x)=\int d^4p \, e^{ipx} \tilde{J}_\Omega(p), \qquad
G^0(x,y)=\int d^4k \, \frac{e^{-ik(x-y)}}{k^2} .
\end{equation}
The right hand side of (\ref{e8}) can be rewritten as
\begin{equation}
J_\Omega(x) G^0(x,y)=\int d^4p \int d^4k \, e^{ipx} \tilde{J}_\Omega(p)
\frac{e^{-ik(x-y)}}{k^2} =\int d^4k \int d^4p \, e^{iky} e^{ipx} 
\frac{\tilde{J}_\Omega(p+k)}{k^2}
\end{equation}
and we obtain the following algebraic equation for the double Fourier
transform of the first order correction to the propagator:
\begin{equation}
p^2 \tilde{G}'(p,k)=\frac{\tilde{J}_\Omega(p+k)}{k^2}.
\end{equation}
Transforming back, in conclusion we find
\begin{equation}
G'(x,y)=\int d^4p \int d^4k \, e^{ipx} e^{iky} \,
\frac{\tilde{J}_\Omega(p+k)}{k^2 p^2}.
\label{e13}
\end{equation}
Therefore, if we know the Fourier transform of the characteristic
function $J_\Omega$ of the spacetime region where the constraint is
imposed, we can in principle compute the leading order correction to
the propagator of the field and thus to $W[J]$.

Now, it is known [3] that the vacuum-to-vacuum amplitude $W[J]=\langle 0^+ 
| 0^- \rangle_J$ of a field system in the presence of an external source $J$ 
is related to the logarithm of the energy of the ground state of the system: 
\begin{equation}
E_0[J]= - T^{-1} \ln W[J],
\label{e2}
\end{equation}
where the functional integral is supposed to be suitably normalized and
the source vanishes outside the temporal interval $[-T/2,+T/2]$, with 
$T$ eventually approaching infinity. (We use units in which $\hbar=c=1$.)

By imposing that the field vanishes in certain compact regions $\Omega_i$ 
of arbitrary shape and position, and taking at the end the derivative of 
$W[J]$ with respect to their relative coordinates, one can find in 
principle the forces present, due to vacuum fluctuations, in a system 
of generalized "uncharged conductors".

A more interesting application of (\ref{e13}) occurs in the case when the
field $\phi(x)$ also interacts with $N$ static pointlike sources placed at
${\bf x}_1,{\bf x}_2...{\bf x}_N$. Namely, let us
add a further, linear coupling term $S_Q$ to the action of the system:
\begin{equation}
S_Q=\int d^4x \, Q(x)\phi(x), \qquad {\rm with} \qquad
Q(x)=\sum_{j=1}^N q_j \delta^3({\bf x}-{\bf x}_j) .
\end{equation}

The energy of the ground state of the system corresponds, up to a constant, 
to the static potential energy of the interaction of the sources through the 
field $\phi$. It is obtained, as before, from the functional average of 
the interaction term, computed keeping the constraint into account:
\begin{equation}
E_0[J,Q] = U({\bf x}_1,...,{\bf x}_N)= 
-T^{-1} \ln \langle \exp \left\{-S_Q\right\} \rangle_J .
\label{e15}
\end{equation}
Expanding (\ref{e15}), one finds that to leading order in the 
$q_j$s, $U({\bf x}_1,...,{\bf x}_N)$ is given by a sum of
propagators integrated on time:
\begin{equation}
U({\bf x}_1,...,{\bf x}_N)= -T^{-1} \sum_{j,l=1}^N q_j q_l
\int dt_j \int dt_l \langle \phi(t_j,{\bf x}_j) \phi(t_l,{\bf x}_l) 
\rangle_J \label{e16}
\end{equation}
where $t_j,t_l \, \epsilon \, [-T/2,+T/2]$. Since the regions $\Omega_i$ are
infinitely elongated in the temporal direction, the function
$\tilde{J}_\Omega(p+k)$ gets factorized as
\begin{equation}
\tilde{J}_\Omega(p+k)=(2\pi)^4\delta(p_0+k_0) \tilde{j}_\Omega({\bf p}+{\bf 
k}). \label{e17}
\end{equation}

Clearly the potential is disturbed by the presence of the "barriers"
$j_\Omega({\bf x})$. We can write, to first order in $\gamma$,
\begin{equation}
U({\bf x}_1,...,{\bf x}_N)=U^0({\bf x}_1,...,{\bf x}_N)+
\gamma U'({\bf x}_1,...,{\bf x}_N).
\end{equation}
and taking into account eq.s (\ref{e13}), (\ref{e17}), we find 
\begin{eqnarray}
U'({\bf x}_1,...,{\bf x}_N) & = & - T^{-1} \sum_{j,l=1}^N q_j q_l
\int dt_j \int dt_l \, G'(x_j,x_l) = \nonumber \\
& = & -(2\pi)^4 T^{-1} \sum_{j,l=1}^N q_j q_l \int dt_j \int dt_l 
\int d^4p \int d^4k \, \frac{e^{ipx_j+ikx_l} \tilde{J}_\Omega(p+k)}{k^2 p^2}= 
\nonumber \\
& = & -(2\pi)^8 T^{-1} \sum_{j,l=1}^N q_j q_l \int dt_j \int dt_l 
\int d^4p \int d{\bf k} \, \frac{e^{ip_0(t_j-t_l)+i{\bf px}_j + i{\bf kx}_l}
\tilde{j}_\Omega({\bf p}+{\bf k})}{(p_0^2+ {\bf k}^2)(p_0^2+ {\bf p^2})}.
\nonumber
\end{eqnarray}
Changing variables to $t=t_j-t_l$ and $s=t_j+t_l$ and integrating,
we finally obtain the 
contribution from the perturbation to the static potential energy:
\begin{equation}
U'({\bf x}_1,...,{\bf x}_N)= -(2\pi)^8
\sum_{j,l=1}^N q_j q_l \int d{\bf p} \int d{\bf k} \,
e^{i{\bf px}_j + i{\bf kx}_l} \, \frac{\tilde{j}_\Omega({\bf p}+{\bf k})}
{{\bf k}^2 {\bf p}^2}.
\label{e19}
\end{equation}

The six-dimensional integral (\ref{e19}), although formally simple, is in
general impossible to evaluate analytically, because the Fourier transform
of the constraint $j_\Omega$ depends in a non trivial way on all angles,
not just on $|{\bf k}|$ and $|{\bf p}|$. In several cases, however, it is
possible to rescale the integration variables dividing them by the sizes
of the regions $\Omega_i$ and by the distances between
the points ${\bf x}_j$ and ${\bf x}_l$.  In this way, any
power-like dependence of $U'$ on the various lengths involved is
factorized in front of the integral; the weak residual dependence (usually
logarithmic) can be estimated by computing the rescaled integral
numerically. 

In order to illustrate this point, let us consider now a special choice
for $j_\Omega$. Clearly, if $j_\Omega$ was exactly equal to the
characteristic function of the set $\Omega$ (i.e., a step function in
coordinates space like above: $j_\Omega({\bf x})=1$ for ${\bf x} \,
\epsilon \, \Omega$, $j_\Omega({\bf x})=0$ otherwise), then the Fourier
transform $\tilde{j}_\Omega({\bf p})$ would have a strongly oscillating
behaviour. To make the integral (\ref{e19}) more accessible to a numerical
computation, it is useful to regularize the step of $j_\Omega({\bf x})$
and "smooth" it. As a further simplification, let us consider a gaussian
function: 
\begin{equation}
j_\Omega({\bf x}) = \exp \left( -\frac{x^2}{a^2}- \frac{y^2}{b^2}- 
\frac{z^2}{c^2} \right) ;
\label{e20}
\end{equation}
this represents a single region $\Omega$, centered at the origin, with 
approximate sizes
$a$, $b$ and $c$ in the three directions. The transform of (\ref{e20}) is
\begin{equation}
\tilde{j}_\Omega({\bf p}) = \pi^{3/2}(abc) \exp \left[ \frac{1}{4}
\left( -a^2 p_x^2- b^2 p_y^2- c^2 p_z^2 \right) \right].
\end{equation}
The generalization to the case of several regions $\Omega_i$, centered
at the positions ${\bf X}_i$, is obvious:
\begin{equation}
\tilde{j}_\Omega({\bf p}) = \sum_i \pi^{3/2}(a_ib_ic_i) 
e^{i{\bf pX}_i} \exp \left[ \frac{1}{4}
\left( -a^2 p_x^2- b^2 p_y^2- c^2 p_z^2 \right) \right].
\end{equation}

After inserting (\ref{e20}) into eq.\ (\ref{e19}),
the integration variables can be rescaled in such a 
way that the integral contains only the ratios between $a$, $b$ and $c$. 
The dependence of the integral on these ratios is quite weak; 
the main dependence is given by the factor $(abc)$ and by the factors 
coming from the re-definition of the integration variables. More detailed 
numerical results will be presented elsewhere.

We observe that the representation of the constraint on the field through
a "smoothed" $j_\Omega({\bf x})$ modifies in part the original physical 
interpretation of our formalism. Instead of perfect conductors, we are 
representing here more realistically some smooth potential barriers in the 
functional integral for the field $\phi$. The probability that the field 
survives, by tunneling, up over the barriers, depends on the parameter 
$\gamma$. 

We have seen that the constraint produces only a small perturbation in the
interaction energy of pointlike sources. Thus our representation of
the constraint by a potential barrier is suitable to describe
non-perfect "conductors", or conductors whose size or thickness is much
smaller than the wavelength characteristic of the interaction between
the pointlike sources.

\bigskip

\noindent
{\bf References.}

\medskip
\noindent
[1] H.B.G.\ Casimir, Proc.\ Kon.\ Ned.\ Akad.\ Wetenschap.\ {\bf
B 51} (1948) 793; C.\ Itzykson and J.-B.\ Zuber, "Quantum Field Theory",
McGraw-Hill, New York, 1980. 

\medskip
\noindent
[2] See for instance A.A.\ Bytsenko, G.\ Cognola, L.\ Vanzo and S.\ 
Zerbini, Phys.\ Rep.\ {\bf 266} (1996) 1.

\noindent
[3] K.\ Symanzik, Comm.\ Math.\ Phys.\ {\bf 16} (1970) 48.
See also G.\ Modanese, Nucl.\ Phys.\ {\bf B 434} (1995) 697.

\end{document}